\documentclass[12pt]{iopart}
\usepackage{graphicx,epsfig}

\def\bea{\begin{equation}}
\def\eea{\end{equation}}
\def\cO{\cos{\O}}
\def\sO{\sin{\O}}
\def\cph{\cos \hat{\varphi}}
\def\sph{\sin \hat{\varphi}}
\def\riff{$K^{'}\{X, \ Y,\ Z\}$}
\def\rif{$K\{x, \ y,\ z\}$}
\def\km{\left(\matrix{\cO & -\sO\ci & \sO\si
\cr \sO & \cO\ci & -\cO\si \cr 0 & \si & \ci \cr}\right)}
\def\gco#1{\rp{GJ}{c^2 \rho^{#1}}}

\def\mash{Mashhoon}
\def\sf{\sin{f}}

\def\ir{{\bf{i}}_{R}}
\def\itt{{\bf{i}}_{T}}
\def\in{{\bf{i}}_{N}}

\def\rfr#1{eq. (\ref{#1})}
\def\rfrs#1#2{eqs. (\ref{#1})-(\ref{#2})}

\def\eqi{\begin{equation}}
\def\eqf{\end{equation}}
\def\eqia{\begin{eqnarray}}
\def\eqfa{\end{eqnarray}}
\def\ci{\cos{i}}
\def\si{\sin{i}}

\def\rp#1#2{{#1\over#2}}

\def\ct#1{\cite{#1}}
\def\lb#1{\label{#1}}

\def\grc{gravitomagnetic}

\def\a{\alpha}

\def\et{\eta}

\def\m{\mu}

\def\p{\pi}

\def\f{\phi}

\def\G{\Gamma}

\def\P{\Pi}

\def\O{\Omega}

\begin{document}

\title{An alternative derivation of the  gravitomagnetic clock
effect}

\author{Lorenzo Iorio\dag, Herbert Lichtenegger\ddag, Bahram Mashhoon\S}

\address{\dag\ Dipartimento di Fisica dell' Universit{\`{a}} di
Bari, via Amendola 173, 70126, Bari, Italy}

\address{\ddag\ Institut f\"{u}r Weltraumforschung,
\"{O}sterreichische Akademie der Wissenschaften, A-8042 Graz,
Austria}

\address{\S\ Department of Physics and Astronomy, University of
Missouri-Columbia, Columbia, Missouri 65211, USA}

\begin{abstract}
The possibility of detecting the gravitomagnetic clock effect
using artificial Earth satellites provides the incentive to
develop a more intuitive approach to its derivation. We first
consider two test electric charges moving on the same circular
orbit but in opposite directions in orthogonal electric and
magnetic fields and show that the particles take different times
in describing a full orbit.  The expression for the time
difference is completely analogous to that of the general
relativistic \grc\ clock effect in the weak-field and slow-motion
approximation. The latter is obtained by considering the
gravitomagnetic force as a small classical non-central
perturbation of the main central Newtonian monopole force. A
general expression for the clock effect is given for a spherical
orbit with an arbitrary inclination angle. This formula differs
from the result of the general relativistic calculations by terms
of order $c^{-4}$.
\end{abstract}

\submitto{\CQG} \pacs{0480}

\section{Introduction} The general relativistic clock effect,
as worked out in \ct{1,2}, consists in the difference in the
orbital periods of two clocks moving in opposite directions along
a circular equatorial orbit around a central rotating mass.  It is
not an easy task to derive the general effect within the framework
of Einstein's theory of gravitation; in fact, for orbits of
arbitrary inclination to the equatorial plane, the clock effect
has been derived only for the case that the orbiting bodies
describe spherical orbits of constant ``radius'' \ct{2}. Moreover,
the case of elliptical orbits has not yet been investigated.

The purpose of this paper is to provide an alternative derivation of this effect for
orbits with zero eccentricity, in order to make its comprehension more intuitive by
stressing and elucidating the analogies and differences between the weak-field and
slow-motion approximation of general relativity and electromagnetism.
General expressions are given for arbitrary values of inclination as well.
They are useful in view of the recent efforts devoted to exploring
the possibility of measuring the clock effect by means of
artificial near-Earth satellites \ct{3,4,5,6,7}.

The paper is organized as follows: In Section 2 we deal with an
electromagnetic system consisting of two test charges orbiting in
opposite directions, acted upon by a central electric field and by
a weaker orthogonal magnetic field.  The latter is treated
perturbatively to first order.  Radiative and
$\mathcal{O}(v^2/c^2)$ effects in such a system are neglected in
order to outline the main features common to the gravitational
case that is treated in Section 3. In this work the
gravitomagnetic force is viewed classically as a non-central small
linear perturbation of the main central Newtonian gravitoelectric
monopole force. Section 4 contains concluding remarks.
\section{The electromagnetic scenario} Let us consider two identical point charges
$q$ of mass $m$ orbiting a central spherically symmetric
distribution of total charge $Q$ of opposite sign, e.g. $Q<0$ and
$q>0$. We suppose that the two charges follow identical but
opposite closed circular orbits and denote the speeds of the
counterclockwise and clockwise moving charges by $v_{+}$ and
$v_{-}$, respectively, i.e. we assume the counterclockwise
direction to be positive. In cylindrical coordinates
$\{\rho,\f\}$, the equation of motion of the two charges reads
\eqi m\rp{v_{\pm}^2}{\rho}=|q|\;E,\eqf with $E=|Q|/\rho^{2}$ and
both charges describe a complete orbit in the same time $T^{(0)}$
\eqi
T^{(0)}=\rp{2\p\rho}{v_{\pm}}=\rp{2\pi}{n}=2\p\sqrt{\rp{m\rho}{|q|\;E}}=
2\p\sqrt{\rp{m\rho^{3}}{|qQ|}},\lb{pere}\eqf where it should be
noted that the mean motion (i.e. orbital frequency)
$n=\sqrt{|q|\;E/m\rho}$ depends on the charge-to-mass ratio $q/m$.

If we switch on a magnetic field ${\bf B}=B{\bf i_z}$ orthogonal
to the plane of motion, the two charges will experience an
additional Lorentz force ${\bf F}_{L}=(q/c){\bf v}\times {\bf B}$,
which for $q>0$ will be antiparallel to the electric field for the
counterclockwise moving charge and parallel to the electric field
for the clockwise moving charge. Hence the equation of motion for
$q>0$ becomes \eqi m\rp{v_{\pm}^2}{\rho}= qE \mp
\rp{q}{c}v_{\pm}B,\lb{mot}\eqf and therefore \eqi
\left(v_{\pm}\pm\rp{qB}{2mc}\rho\right)^2 = \left(n^2 +
\rp{q^2B^2}{4m^2c^2}\right)\rho^2.\lb{vel2}\eqf

We assume that the magnetic field is weak; therefore, we can
neglect the square of the Larmor frequency ${q^2B^2}/{4m^2c^2}$ in
comparison with the square of the orbital frequency $n^{2}$ in
\rfr{vel2} and obtain $v_{\pm}=(n\mp qB/2mc)\rho$, i.e. \eqi
\left({\rp{d\f}{dt}}\right)_{\pm}=\pm\left(n\mp
\rp{qB}{2mc}\right),\lb{dft}\eqf or \eqi dt_{\pm}=\pm\rp{d\f}{ n\
(1\mp \rp{qB}{2mcn})}\simeq
\pm\rp{d\f}{n}\left(1\pm\rp{qB}{2mcn}\right).\lb{piuomen}\eqf By
integrating \rfr{piuomen} from 0 to $2\p$ for the counterclockwise
orbit and $2\p$ to 0 for the clockwise orbit, we find the orbital
periods of the two charges \eqi
T_{\pm}=\frac{2\pi}{n}\left(1\pm\frac{qB}{2mcn}\right)=T^{(0)}\pm
\rp{\p}{c}\rp{B}{E}\rho,\lb{mcp}\eqf and hence their difference
after one revolution \eqi
T_{+}-T_{-}=\rp{2\p}{c}\rp{B}{E}\rho.\lb{dt}\eqf

By inspection of \rfr{mcp} we see that the magnetic correction to the orbital period is
independent of the charge-to-mass ratio of the orbiting electric charges, in contrast to
the unperturbed period $T^{(0)}$.  Indeed we could have started Section 2 with the less
restrictive assumption that the two charges only have the same charge-to-mass ratio.

It is interesting to observe that \rfr{dt} is an exact consequence
of \rfr{mot}, i.e. the weak-field approximation is unnecessary for
the validity of this relation. This circumstance has an analogue
in the gravitational case discussed in Section 3; that is,
\rfr{hjk} below turns out to be exact for circular orbits in the
equatorial plane of the exterior Kerr spacetime.

Let us now assume that the two charges are far away from the
central charge and current distributions so that the magnetic
field can be considered to be generated by a magnetic dipole ${\bf
m}=-\m{\bf i_z}$ of magnitude $\m=IS/c$, where $S$ is the surface
area of the loop enclosed by the current $I$ and therefore \eqi
B=\rp{\m}{\rho^3},\hspace{5mm}E=\rp{|Q|}{\rho^2}\lb{ele},\eqf
which, upon inserting into \rfr{dt} yield \eqi
T_{+}-T_{-}=\frac{2\p}{c}\rp{\m}{|Q|}.\lb{dtel} \eqf This time
difference depends on both the sign of the charges and on the
direction of the magnetic field. Upon exchanging the signs of $Q$
and $q$, i.e. $Q>0$ and $q<0$, the counterclockwise revolving
charge will move faster while the clockwise moving charge will
move slower. However, as expected by charge symmetry, \rfr{dtel}
will be unaffected if the signs of the charges and of the magnetic
field are reversed simultaneously. Further, we note that the
radius of the orbit does not appear in \rfr{dtel} and that this
time difference can be interpreted as a consequence of the fact
that the speed of light has a  finite value; indeed, for
$c\rightarrow\infty$, $T_{+}-T_{-}\rightarrow 0$.

The main considerations of this section are related, via the
Larmor theorem, to certain interesting phenomena in rotating
frames of reference \ct{8,9}.

Orbits of charged particles off the equatorial plane are no longer
spatially closed in general because their instantaneous planes
undergo Larmor precession induced by the magnetic  field and it
will thus be necessary to define the relevant periods $T_{\pm}$ in
terms of azimuthal closure involving a complete loop in the $\phi$
coordinate.  For the description of such a configuration it is
useful to introduce the local frame attached to the moving
particle, where $\ir,\ \itt,\ \in$ denote its orthogonal unit
vectors related to the radial, along-track and cross-track
directions, respectively. In fact, $\itt$ denotes the orthogonal
direction, in the instantaneous orbital plane, to the radial one;
in general, it does not coincide with the direction along the
track unless one considers circular orbits.  We will return to the
case of spherical orbits in Section 4, where the local frame
described here is employed along the unperturbed orbit.

\section{The gravitational scenario} The electromagnetic scenario previously described is
analogous to the following gravitational one: Let us  consider a
central spherically symmetric mass $M$ rotating slowly with its
proper angular momentum directed along the $z$-axis, ${\bf
J}=J{\bf i_z}$, of an asymptotically inertial frame \rif\, whose
$(x,y)$-plane coincides with the equatorial plane of the
gravitating source, and a pair of test bodies orbiting along a
circular equatorial path in opposite directions. Further, we
assume that the radius $\rho$ of the orbits is much larger than
the Schwarzschild radius $r_g = 2GM/c^2$ of the central body; this
would apply to an experiment in the field of the Earth, for
instance. It is well known that in the weak-field and slow-motion
approximation of general relativity the stationary spacetime
metric of a rotating ``spherically symmetric'' mass-energy
distribution generates the so-called gravitoelectric and \grc\
fields \ct{10,11} \eqi{\bf E}_g=-\rp{GM}{r^2}{\bf i}_R\lb{eg},\eqf
\eqi{\bf B}_g={\mathcal{r}}\times{\bf
A}_g=\rp{2G}{c}\left[\rp{{\bf J}-3({\bf J}\cdot{\bf i}_R){\bf
i}_R}{r^3}\right]\;,\lb{opl}\eqf with the \grc\ potential given by
\eqi{\bf A}_g=-\rp{2G}{c}\rp{{\bf J}\times{\bf r}}{r^3}.\eqf  At
this point the calculations follow closely those of the
electromagnetic case previously examined in Section 2 because the
equation of motion of a test body in the ``weak'' gravitational
field of a general stationary axisymmetric mass-energy
distribution is analogous to that of a point charge $q$ acted upon
by electric and magnetic fields {\bf E} and {\bf B}, \eqi m{\bf
a}_{g}=m({\bf E}_{g}+\rp{\bf v}{\it c}\times{\bf
B}_{g}).\lb{etrm}\eqf Therefore, by reasoning as in the
electromagnetic case for circular orbits with zero inclination
(Section 2), we obtain \eqi T_{+}-T_{-}=\rp{2\p}{c}\rp{B_{g}}{
E_{g}}\rho.\lb{dyt}\eqf


For equatorial circular orbits, eq. (\ref{opl}) immediately yields
\eqi{\bf B}_g=\rp{2G}{c}\rp{J}{\rho^3}{\bf i_z}.\lb{gcb}\eqf
By inserting $E_g$ and
$B_g$ from eqs. (\ref{eg}) and (\ref{gcb}) into \rfr{dyt}
we obtain the well-known expression
\eqi T_{+}-T_{-}=4\p\rp{J}{c^2 M}.\lb{hjk}\eqf

It is an interesting feature of \rfr{hjk} that the mass moving in the same sense of
rotation as the central mass moves slower than the mass moving in the opposite direction.
If we reversed the sense of rotation of the central gravitating source, the clockwise
moving test mass would be slower. In this way the sense of rotation of the central mass is
no longer a matter of convention but could be related to a physical phenomenon, i.e. the
mass loop moving slower. Also in this case, in the limit $c\rightarrow\infty$,
$T_{+}-T_{-}\rightarrow 0$.

An interesting feature of gravitoelectromagnetism is the
gravitational Larmor theorem \ct{12} according to which
gravitoelectric and gravitomagnetic fields are locally equivalent
to translational and rotational accelerations of an observer in
Minkowski spacetime.  The gravitational Larmor theorem is
therefore in effect Einstein's principle of equivalence in the
gravitoelectromagnetic context. Note that in \rfr{hjk} the
gravitoelectric Keplerian periods
$T^{(0)}=2\pi/n=\sqrt{GM/\rho^{3}}$ cancel out, no matter what the
masses of the orbiting particles are, in accordance with the
equivalence principle and in contrast to the electromagnetic case,
where the unperturbed periods depend on the charge-to-mass ratio
of the particles.

Finally, it is worth noting that the \grc\ correction to the
unperturbed Keplerian period is independent of the radius of the
orbit, a feature also common to the electrodynamic case. Moreover,
the Newtonian constant $G$ does not appear in \rfr{hjk}: this fact
may account for the unexpectedly large value of the time shift in
the field of the Earth which amounts to about $10^{-7}$ s.  In
some sense, this classical effect is a gravitomagnetic analog of
the topological Aharonov-Bohm effect.  It turns out that if we
deal with the {\it proper} periods of the test bodies, i.e. the
periods according to comoving clocks, then the result is the same
as in eq. (17) up to terms of order $c^{-4}$ that depend on $G$
and $\rho$.

\section{Arbitrary inclination: inertial azimuthal closure} If the orbital plane has an arbitrary inclination $i$ to
the equatorial plane of \rif, a similar reasoning as in Section 2
holds: the orbital plane undergoes  Lense-Thirring precession,
which is the gravitational analog of the magnetic Larmor
precession.  Let us imagine that in the absence of the
gravitomagnetic field, the orbit is a circle of radius $r$ in a
fixed plane that is inclined with respect to the equatorial
$(x,y)$-plane by the inclination angle $i$; that is, the normal to
the orbital plane is tilted away from the $z$-axis by the angle
$i$. Moreover, the longitude of the ascending node is given by the
azimuthal angle $\Omega$, as in Figure 1.  Once the
gravitomagnetic field is ``turned on,'' the orbit will twist out
of this fixed plane.

 In order to derive analytically the time
$T$ needed to pass from $\phi_0$ at $t_0=0$ to $\phi_0+2\pi$ at
$t_0+T$ in the equatorial plane of the inertial observer we will
use the following reference frames: the asymptotically inertial
frame $K\{x,\ y,\ z\}$ previously defined and a frame $K^{'}\{X, \
Y,\ Z\}$ with the $Z$-axis directed along the orbital angular
momentum ${\bf L}$ of the unperturbed test body, where the $(X,
Y)$-plane coincides with the unperturbed orbital plane of the test
particle and the $X$-axis is directed along the line of nodes.
$K\{x,\ y,\ z\}$ and $K^{'} \{X, \ Y,\ Z\}$ have the same origin
located at the center-of-mass of the central body. The
transformation matrix ${\bf R}_{xX}$ for the change of coordinates
from $K^{'} \{X, \ Y,\ Z\}$ to $K\{x,\ y,\ z\}$ is given by \eqi
{\bf R}_{xX}=\km \lb{mat}.\eqf Our calculation for the azimuthal
period $T$ will be valid for a counterclockwise orbit as in Figure
1; however, it is clear from the symmetry of the configuration
that the result for the clockwise case ($T_{-}$) can be simply
obtained from our result by reversing the sign of the perturbation
term in $T_{+}$. In the asymptotically inertial frame \riff, let
us choose cylindrical coordinates $\{\rho,\ \varphi,\ Z\}$ and
write \rfr{etrm} in these coordinates. The unperturbed orbit is
given by $\rho\equiv r=$ constant, $Z=0$ and
$\varphi=\varphi_0+nt$ (with $n=\sqrt{GM/\rho^{3}}$). Since we
will consider the gravitomagnetic acceleration
$c^{-1}\textbf{v}\times\textbf{B}_g$ as a small perturbation of
the main gravitoelectric monopole term $\textbf{E}_g$, we evaluate
the disturbing acceleration with respect to the unperturbed orbit.
By using eqs. (\ref{eg}) and (\ref{opl}), \rfr{etrm} can be
written as

\begin{eqnarray} a_{\rho}=\ddot{\rho}-\rho\
{\dot{\varphi}}^2=-\rp{GM}{\rho^2}+2\gco{2}n\ci\lb{grc1},\\
 a_{\varphi}=\rho\ \ddot{\varphi}+2\dot{\rho}\ \dot{\varphi}=0\;,\\
a_{Z}=\ddot{Z}=-n^{2}Z+4\gco{2}n\si\sin{(\varphi_0+nt)}.\lb{grc3}
\end{eqnarray}

\noindent Here the radial and along-track components of the
acceleration are given by $a_{\rho}=\ddot{X}{\rm cos}\ \varphi +
\ddot{Y} {\rm sin}\ \varphi$ and $a_{\varphi} = - \ddot{X}{\rm
sin}\ \varphi + \ddot{Y}{\rm cos}\ \varphi,$ respectively. Note
that in \rfr{grc3} we retain $Z$, because due to the
gravitomagnetic non-central acceleration, the motion is no longer
confined to a plane and therefore $Z$ will be proportional to the
gravitomagnetic perturbation and treated to first order.

For a spherical orbit $\rho$ remains constant to first order;
therefore, from \rfr{grc1} we obtain \eqi
\varphi=\varphi_0+nt-\gco{3}t\ci\equiv\hat{\varphi}+\delta\varphi
,\lb{fi}\eqf where $\hat{\varphi}=\varphi_0+nt$. It follows from
the linear dependence of $\varphi$ on $t$ in eq. (22) that eq.
(20) is satisfied. The solution of \rfr{grc3} reads \eqi
Z=-2\gco{2}t\si\cos{(\varphi_0+nt)}+k_1\sin nt+k_2\cos
nt,\lb{zeta}\eqf where $k_1$ and $k_2$ are constants of
integration. If we assume that at $t_0=0$ the perturbed orbit
agrees with the unperturbed orbit, i.e. $Z(0)=0$, then $k_2=0$ and
we will treat $k_1$ to first order of the perturbation in what
follows. Let us write the general solution of \rfrs{grc1}{grc3} as
$X=\rho\cos \varphi(t)$ and $Y=\rho\sin \varphi(t)$ together with
\rfr{zeta}. By using $\cos (\hat{\varphi}+\delta\varphi) \simeq
\cos\hat{\varphi}-\delta\varphi\sin \hat{\varphi}$ and $\sin
(\hat{\varphi}+\delta\varphi) \simeq \sin
\hat{\varphi}+\delta\varphi\cos \hat{\varphi}$, we find \eqi
X=\rho\cos \hat{\varphi}+\gco{2}t\ci\sin\hat{\varphi},\lb{ics}\eqf
\eqi Y=\rho\sin
\hat{\varphi}-\gco{2}t\ci\cos\hat{\varphi},\lb{ipsilon}\eqf \eqi
Z=-2\gco{2}t\si\cos{\hat{\varphi}}+k_1\sin nt.\lb{zetag}\eqf From
these equations we obtain by means of the transformation
(\ref{mat}) the solution in $K\{x,\ y,\ z \}$,
$$x=\rho(\cO\cph-\ci\sO\sph)+k_1\si\sO\sin nt$$
\eqi+
\gco{2}t[\ci\cO\sph+(\cos^{2}i-2\sin^{2}i)\sO\cph],\lb{ICS}\eqf
$$y=\rho(\sO\cph+\ci\cO\sph)-k_1\si\cO\sin nt$$
\eqi+\gco{2}t[\ci\sO\sph-(\cos^{2}i-2\sin^{2}i)\cO\cph],\lb{YPSILON}\eqf
\eqi z=\rho\si\sph+k_1\ci\sin nt-\rp{3}{2}\gco{2}t\sin
2i\cph.\lb{ZETA}\eqf The temporal behavior of the azimuthal angle
$\phi$ can be obtained via \eqi\tan
\phi=\rp{y(t)}{x(t)},\lb{tang}\eqf and the time $T$ needed to pass
from $\phi_0$ at $t_0=0$ to $\phi_0+2\pi$ at $t_0+T$ follows upon
expanding the relation \eqi\tan
\phi_0=\tan(\phi_0+2\pi),\lb{tang2}\eqf where \eqi\tan
\phi_0=\rp{y(0)}{x(0)}=\rp{\sO\cos\varphi_0+\ci\cO\sin\varphi_0}{\cO\cos\varphi_0-\ci\sO\sin\varphi_0}\lb{tang3},\eqf
and \eqi\tan(\phi_0+2\pi)=\rp{y(T)}{x(T)}.\lb{tang4}\eqf Since the
deviation from the unperturbed Kepler period $T^{(0)}=2\pi/n$ will
be small, let us write \eqi
T=\frac{2\pi}{n}(1+\epsilon),\lb{tpert}\eqf with $\epsilon \ll 1$
and further \eqi \sin nT=\sin(2\pi+2\pi\epsilon)\simeq
2\pi\epsilon\lb{memm},\eqf \eqi
\cph(T)=\cos(\varphi_0+nT)=\cos(\varphi_0+2\pi+2\pi\epsilon)\simeq\cos\varphi_0-2\pi\epsilon\sin\varphi_0\lb{comm},\eqf
\eqi
\sph(T)=\sin(\varphi_0+nT)=\sin(\varphi_0+2\pi+2\pi\epsilon)\simeq\sin\varphi_0+2\pi\epsilon\cos\varphi_0.\lb{simm}\eqf
Therefore, in the calculations for $x(T)$ and $y(T)$, terms
proportional to $k_1\sin nT$, due to \rfr{memm}, will be of second
order and will be neglected. Hence we find
$$x(T)=\rho(\cO\cos\varphi_0-\ci\sO\sin\varphi_0)-2\pi\epsilon\rho\;(\cO\sin\varphi_0+\ci\sO\cos\varphi_0)$$
\eqi+
2\pi\gco{2}\rp{1}{n}[\ci\cO\sin\varphi_0+(\cos^{2}i-2\sin^{2}i)\sO\cos\varphi_0]\lb{ICST},\eqf
$$y(T)=\rho(\sO\cos\varphi_0+\ci\cO\sin\varphi_0)-2\pi\epsilon\rho\;(\sO\sin\varphi_0-\ci\cO\cos\varphi_0)$$
\eqi+2\pi\gco{2}\rp{1}{n}[\ci\sO\sin\varphi_0-(\cos^{2}i-2\sin^{2}i)\cO\cos\varphi_0].\lb{YPSILONT}\eqf
Using \rfrs{tang2}{tang4} and \rfrs{ICST}{YPSILONT}, we find after
some algebra \eqi
\epsilon=\gco{3}\rp{\ci}{n}(1-2\tan^{2}i\cos^{2}\varphi_0).\eqf As
expected, $\epsilon$ is proportional to $J$ and vanishes for a
nonrotating source. By means of \rfr{tpert} it finally follows
that \eqi T_{\pm}=\rp{2\pi}{n}[1\pm\gco{3}\rp{\ci}{n}(1-2\
{\tan}^{2}i \cos^{2}\varphi_0)],\lb{TP}\eqf or \eqi
T_{+}-T_{-}=4\pi\rp{J\ci}{c^2 M }(1-2\ {\tan}^{2}i
\cos^{2}\varphi_0).\lb{DTP}\eqf

Let us note that for $i=0$, we recover the result of Section 3. On
the other hand, $\epsilon$ diverges for $i = \pi/2$.  It follows
from general relativity that for a geodesic (spherical) polar
orbit, the clock effect disappears since the angular momentum
vector of the source in effect lies in the orbital plane. The
orbital period is then simply given by the gravitoelectric
Keplerian period.  On the other hand, the period for {\it
azimuthal closure} is given by $2\pi(2GJ/c^2\rho^3)^{-1}$, which
is very long compared to the Keplerian period \ct{13}. This
circumstance is reflected in our first-order perturbative result
given by eq. (41): For
$i\rightarrow\pi/2\;,\;T_{\pm}\rightarrow\infty$. Therefore, in
eq. (41), the inclination angle $i$ must be sufficiently less than
$\pi/2$ such that the perturbative treatment in Section 4 remains
valid.

An important feature of eq. (41) is that when $i\neq0, T_{\pm}$
depends upon $\varphi_0$, i.e. the clock effect depends in general
on the position of the mass $m$ along the orbit at $t=0$.  This
dependence of the clock effect on where the mass $m$ is along the
orbit when the timing observations begin is illustrated in Fig. 2
and could be helpful in the detection of this effect.

Up to now we have assumed that when unperturbed the two satellites
orbit along opposite directions in the same plane with arbitrary
inclination $i$. Let us now consider the case of two masses
$m_{+}$ and $m_{-}$ having the same distance from the center but
moving in different orbital planes, say $0<i_{+}<\pi/2$ and
$\pi/2<i_{-}<\pi$, respectively. Following the same reasoning as
before, their orbital periods, as viewed by a static,
asymptotically inertial observer and after choosing suitably $t_0$
so that $\varphi_0=\pi/2 $, can be written as \eqi T_{-} =
T^{(0)}+2\pi\frac{J}{c^{2}M}\cos{i_{+}}\lb{tpiu},\eqf \eqi T_{-} =
T^{(0)}-2\pi\frac{J}{c^{2}M}\cos{i_{-}}.\lb{tmen} \eqf From these
expressions, we immediately find \eqi T_{-}-T_{-} =
2\pi\frac{J}{c^{2}M}(\cos{i_{+}}+\cos{i_{-}}),\lb{dtpiu}\eqf \eqi
T_{-}+T_{-} =
2T^{(0)}+2\pi\frac{J}{c^{2}M}(\cos{i_{+}}-\cos{i_{-}}).\lb{dtmen}\eqf

From \rfr{dtmen} it is seen that the sum of the orbital periods of
the two point masses will also show a gravitomagnetic contribution
provided the inclinations of the two satellites are different.
Note that \rfr{dtpiu} reduces correctly to \rfr{DTP} if the two
orbital planes coincide, while in \rfr{dtmen} the gravitomagnetic
contribution  vanishes. Moreover, a very interesting feature
arises for supplementary inclinations of the two satellites, i.e.
$i_{+}+i_{-}=180^{\circ}$; indeed, in this case
\rfrs{dtpiu}{dtmen} become \eqi T_{-}-T_{-} = 0\lb{dtpiun},\eqf
\eqi T_{-}+T_{-} =
2T^{(0)}+4\p\rp{J}{c^{2}M}\cos{i_{+}}.\lb{dtmenn}\eqf

One of the most striking implications of \rfr{dtmenn} is that the
LARES mission \ct{14}, originally proposed  to detect the
Lense-Thirring drag of the orbital plane, could also be used to
detect the gravitomagnetic contribution to the sum of the orbital
periods. However, due to the present uncertainty of the value of
$GM_\oplus,\delta (GM_\oplus)=8\times 10^{11}\ {\rm cm}^{3}\ {\rm
s}^{-2}$ \ct{15}, the error in the unperturbed Keplerian period of
the LAGEOS satellite is larger than the effect to be measured \eqi
\delta T^{(0)}=1.52\times 10^{-5}\ {\rm s},\eqf \eqi
4\pi\rp{J}{c^{2}M}\cos i_+ =4.71\times 10^{-8}\ {\rm s}.\eqf
Therefore, an improvement of our knowledge of $GM_\oplus$ would be
necessary before any observables involving the sum of the
unperturbed orbital periods and of the gravitomagnetic corrections
may become detectable.

Finally, these results suggest that the gravitomagnetic clock effect may be enhanced by
considering suitable constellations of satellites orbiting the Earth.

\section{Conclusions} Exploiting the formal analogy between the law of motion of a charged
particle acted upon by an electromagnetic field and the weak-field
and slow-motion approximation of general relativity
(``gravitoelectromagnetism''), it has been possible to derive in a
simple fashion the gravitomagnetic clock effect for a couple of
point masses following spherical orbits in space. General
expressions for arbitrary values of inclination angle to the
equatorial plane are given for these spherical orbits. These
results have been obtained by neglecting terms of
$\mathcal{O}(c^{-4})$; in this way, \rfr{DTP} is equally valid for
proper periods of comoving clocks.
If the two satellites orbit in planes with different inclinations,
the sum of their orbital periods also exhibits a gravitomagnetic
part.

It is worth-while to compare our main result \rfr{DTP} with the
corresponding general relativistic expression (see \ct{2} p. 143,
eqs. (36)-(37)), \eqi T_{+}-T_{-}=4\p\rp{J}{c^2
M}\lambda^{'}\cos{\a},\lb{mash}\eqf where
\eqi\lambda^{'}=\lambda-3\Phi_{0}\G_{0}=\G_{0}-2
\G_{0}^{-1}\tan^{2}\a \cos^{2}\et_0
-3\Phi_{0}\G_{0}.\lb{lambda}\eqf Here, $\a$ and $\et_{0}$ must be
replaced by $i$ and $\varphi_{0}$, respectively, and
$\G_{0}\rightarrow 1$, $\Phi_{0}\rightarrow 0$ provided that terms
of $\mathcal{O}(c^{-4})$ are neglected. For the proper periods of
comoving clocks an equation similar to \rfr{mash} holds, except
that $\lambda^{'}$ must be replaced by $\lambda$. Note that these
results have been obtained in a perturbative way and hold only for
$\alpha$ sufficiently different from $\pi/2$. Thus our approach
gives the same result for $T_{+}-T_{-}$ as general relativity once
terms of order $c^{-4}$ are neglected.\

As it is for all general relativistic effects, the observation of
the gravitomagnetic clock effect is  a tremendously difficult
undertaking as well. This becomes immediately clear by noting that
for a near-Earth orbit ($T\sim 10^4$ s) a time variation of $\sim
100$ ns can be caused equally well by a radial or azimuthal
deviation of $\sim 0.1$ mm from the ideal orbit and therefore all
forces that may produce accelerations larger than $\sim 10^{-12}$
m/s$^2$ must be taken into account. The empirical verification of
the clock effect essentially faces two problems: (a) to measure
with the utmost precision the actual position of the satellites,
(b) to model with extreme accuracy all perturbing forces which
will influence the period of the satellites. For a single orbit,
this goal is certainly unattainable, but it may become feasible
after a sufficiently long time of observation due to the
accumulative character of the clock effect. While the present
satellite-to-satellite tracking techniques allow the determination
of an orbit with an accuracy of $\sim 1$ cm so that a minimum of
$\sim 1000$ revolutions will be needed for the clock effect to
become detectable, the consideration of the perturbing forces at
the required level is quite demanding. Non-gravitational
perturbations may be overcome by means of modern drag-free
technology; however, the correct determination of all
gravitational effects is limited by the accuracy of the respective
Earth gravity field models that are currently available.
Preliminary results suggest that it is in particular the
uncertainty in the even zonal harmonics of the spherical expansion
of the terrestrial gravitational field \ct{7} as well as the zonal
tidal perturbations \ct{4} that presently inhibits the successful
realization of the clock experiment.  It should be mentioned,
however, that upcoming geodetic space missions (especially GRACE
and GOCE) are expected to improve significantly the accuracy of
the gravity field of the Earth and may then allow the observation
of the gravitomagnetic clock effect within a few percent accuracy.

\ack L. Iorio is grateful to L. Guerriero and I. Ciufolini for
their support and encouragement.


\begin{figure}
\begin{center}
\epsfbox{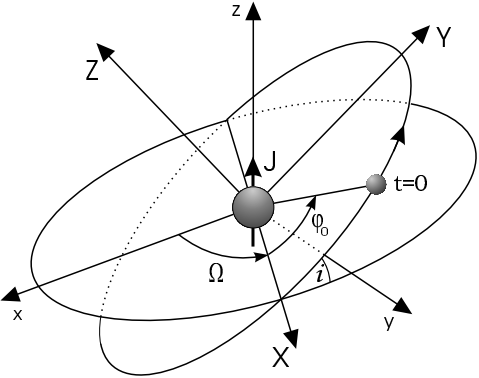}
\end{center}
\caption{\label{figure1} Schematic diagram of the unperturbed
circular trajectory for the general spherical orbit. In practice,
the instantaneous orbital plane at the instant observations begin,
i.e. $t=0$, can be taken to be the plane of the unperturbed
orbit.}
\end{figure}


\begin{figure}
\begin{center}
\epsfbox{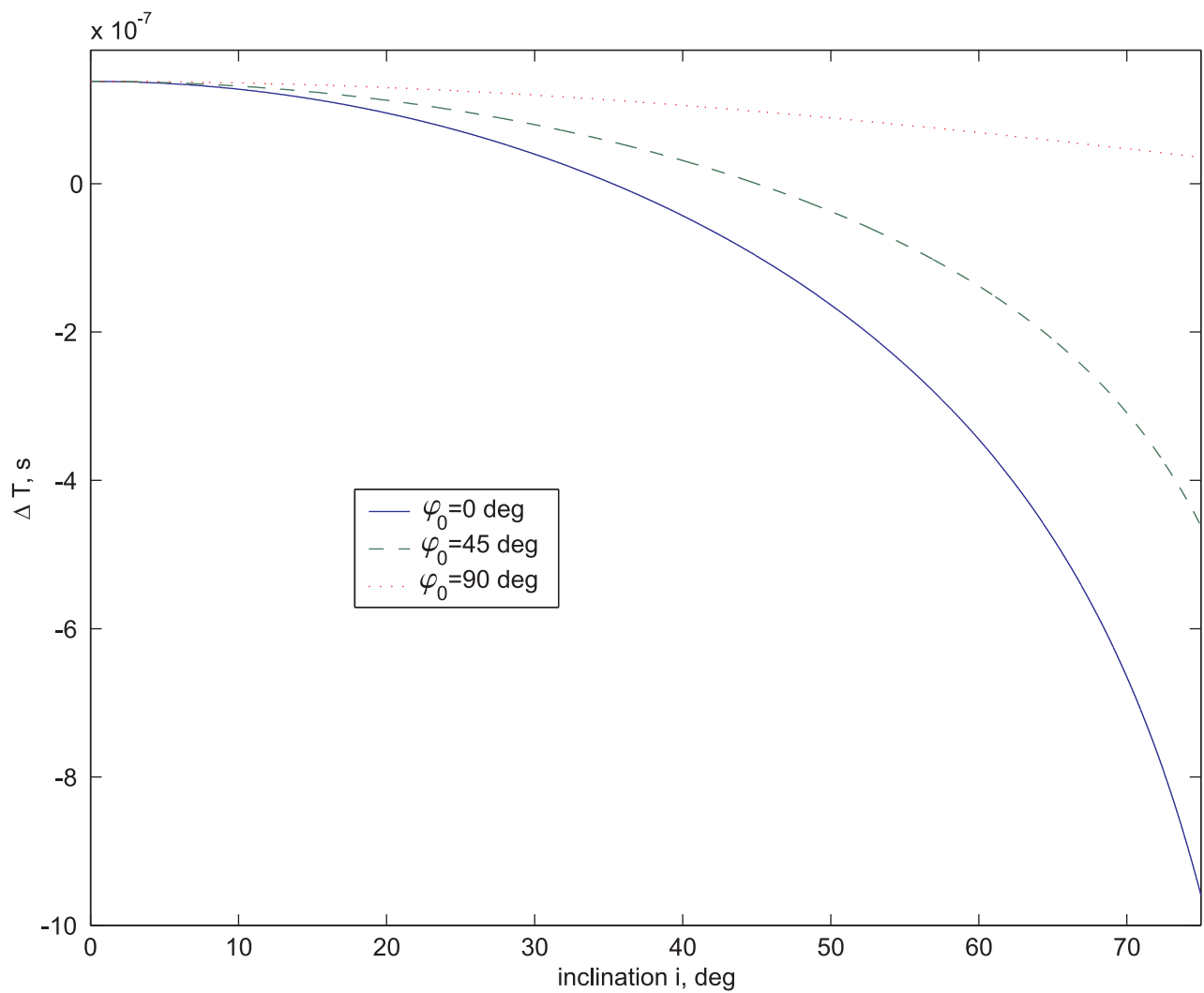}
\end{center}
\caption{\label{figure2} Plot of $\Delta T = T_{+} - T_{-}$, given
in eq. (42), in seconds versus the orbital inclination $i$ in
degrees for a satellite in a spherical orbit around the Earth. For
inclined orbits, the three graphs illustrate the dependence of the
gravitomagnetic clock effect on $\varphi_0$, which is the angular
position of the satellite along its circular orbit at $t$ = 0
measured from the line of the ascending node (cf. Figure 1).}
\end{figure}


\end{document}